\documentclass[a4paper]{jpconf}
\usepackage{graphicx}
\usepackage{amsmath}

\newcommand{\V}[1]{\mathbf{#1}}
\newcommand{\degree}{\ensuremath{^\circ }}

\begin{document}

\addcontentsline{toc}{chapter}{Bibliografia}
\bibliographystyle{unsrt}

\setcounter{page}{1}
\pagenumbering{arabic}

\begin{center}
Accepted for publication on Advances in Astronomy
\end{center}
\vspace{-0.5cm}

\title{Latitudinal Dependence of Cosmic Rays Modulation at 1 AU and Interplanetary-Magnetic-Field Polar Correction}

\author{P Bobik$^1$, G Boella$^{2}$, MJ Boschini$^{2,3}$, C Consolandi$^{2,4}$, S Della Torre$^{2,5}$,
         M Gervasi$^{2,4}$, D Grandi$^2$, K Kudela$^1$, S Pensotti$^{2,4}$, PG Rancoita$^2$,
        D Rozza$^{2,5}$, M Tacconi$^{2,4}$}

\address{
$^1$ Institute of Experimental Physics, Kosice (Slovak Republic)\\
$^2$ INFN Milano-Bicocca, Milano (Italy)\\
$^3$ CILEA, Segrate (Milano, Italy)\\
$^4$ University of Milano-Bicocca, Milano (Italy)\\
$^5$ University of Insubria, Como (Italy)}



\ead{piergiorgio.rancoita@mib.infn.it}

\begin{abstract}
The cosmic rays differential intensity inside the heliosphere,
for energy below 30 GeV/nuc, depends on solar activity
and interplanetary magnetic field polarity. This variation, termed solar modulation,
is described using a 2-D
(radius and colatitude) Monte Carlo approach for solving the Parker transport equation that includes diffusion,
convection, magnetic drift and adiabatic energy loss.
Since the whole transport is strongly related to the interplanetary magnetic field (IMF) structure,
a better understanding of his description is needed in order to reproduce the cosmic rays intensity at the Earth,
as well as outside the ecliptic plane.
In this work an interplanetary magnetic field model including the standard description on ecliptic region and
a polar correction is presented. This treatment of the IMF, implemented in
the HelMod Monte Carlo code (version 2.0), was used to determine the effects on the differential intensity of Proton at 1\,AU and allowed one to investigate how
latitudinal gradients of proton intensities, observed in the inner heliosphere with the
Ulysses  spacecraft during 1995, can be affected by the modification of the IMF in the polar regions.
\end{abstract}

\section{Introduction}
The Solar Modulation, due to the solar activity, affects
the Local Interstellar Spectrum (LIS) of Galactic Cosmic Rays (GCR) typically at energies lower than 30 GeV/nucl.
This process, described by means of the Parker equation~(e.g., see~\cite{parker1965,Bobik2011ApJ}
and Chapter 4 of~\cite{rancoita2011}), is originated from the interaction of GCRs with the
interplanetary magnetic field (IMF) and its irregularities.
The IMF is the magnetic field that is carried outwards during the solar wind expansion.
The interplanetary conditions vary as a function of the
solar cycle which approximately lasts eleven years.
In a solar cycle, when the maximum activity occurs, the IMF reverse his polarity.
Thus, similar solar polarity conditions are found almost every 22 years~\cite{strauss2012}.
In the HelMod Monte Carlo code version 1.5 (e.g., see Ref. \cite{Bobik2011ApJ}), the ``classical'' description of IMF, as proposed by Parker~\cite{parker58},
was implemented together with the polar corrections of the solar magnetic field suggested subsequently in~\cite{JokipiiKota89,langner2004}.
This IMF was used inside the HelMod~\cite{Bobik2011ApJ} code to investigate the
solar modulation observed at Earth and to partially account for GCR latitudinal gradients, i.e., those observed with
the Ulysses spacecraft~\cite{Heber1996,Simpson1996a}.
In order to fully account for both the latitudinal gradients and latitudinal position of the proton-intensity minimum observed during the Ulysses fast scan in 1995, the HelMod Code was updated to the version 2.0 to include a new
treatment of the parallel and diffusion coefficients following that one described in Ref.~\cite{Strauss2011}.
In the present formulation, the parallel component of the diffusion
tensor depends only on the radial distance from the Sun, while it
is independent of solar latitude.

\section{The Interplanetary Magnetic Field}\label{IMF}

Nowadays, we know that there is a Solar Wind plasma (SW) that permeates the
interplanetary space and constitutes the interplanetary medium.
In IMF models the magnetic-field lines are supposed
to be embedded in the non-relativistic streaming particles of the SW, which carries the field with them into interplanetary
space,  producing the large scale structure of the IMF and the heliosphere.
The ``classical'' description of the IMF was proposed originally by Parker~(e.g., see \cite{Bobik2011ApJ,parker58,parker1957,Parker1960,Parker1961a,Parker1963} and Chapter 4 of~\cite{rancoita2011}).
He assumed i) a constant solar rotation with angular velocity ($\omega$),
ii) a simple spherically symmetric emission of the SW
and iii) a constant (or approaching an almost constant) SW speed ($V_{\textrm{sw}}$) at larger
radial distances ($r$), e.g., for $r > r_b \approx 10 $R$_\odot$ (where R$_\odot$ is the Solar radius), since beyond $r_b$
the wind speed varies
slowly with the distance.
The ``classical'' IMF can be analytically expressed as~\cite{Hattingh1995}
\begin{equation}\label{eq::parkerBP}
 \V{B}_{Par} = 	\frac{A}{r^2}(\V e_r - \Gamma\V e_\varphi)[1-2H(\theta-\theta')],
\end{equation}
where $A$ is a coefficient that determines the IMF polarity and allows
$|\V B_{Par} |$ to be equal to $B_\oplus$, i.e.,
the value of the IMF at Earth's orbit as extracted from NASA/GSFC's OMNI data set through OMNIWeb
\cite{SW_web,OMNIWeb};  $\V e_r $ and $\V e_\varphi $ are unit vector components in the
radial and azimuthal directions, respectively; $\theta$ is the colatitude (polar angle);
$\theta'$ is the polar angle determining the position of
the Heliospheric Current Sheet (HCS)\cite{JokipiiThomas1981};
$H$ is the Heaviside function: thus, $[1-2H(\theta-\theta')]$ allows $\V B_{Par}$
to change sign in the two regions above and below the HCS \cite{JokipiiThomas1981} of the heliosphere; finally,
\begin{equation}
 \Gamma=\tan\Psi=\frac{\omega (r-r_b) \sin\theta}{V_{\textrm{sw}}}
\end{equation}
with $\Psi$  the spiral angle. In the present model $\omega$ is assumed to
be independent of the heliographic latitude and equal to the
sidereal rotation at the Sun's equator.
The magnitude of Parker field is thus:
\begin{equation}\label{eq::parkerBP_mag}
B_{Par} = 	\frac{A}{r^2}\sqrt{1+ \Gamma^2}.
\end{equation}
\begin{table}[b]
\begin{center}
 \begin{tabular}{lll}
  \hline
  &  Period         &  Years\\
  \hline\hline
  I)   & $A<0$ Ascending  & 1964.79--1968.87, 1986.70--1989.54, 2008.95--2009.95\\
  II)  & $A<0$ Descending & 1964.53--1964.79, 1979.95--1986.70, 2000.28--2008.95\\
  III) & $A>0$ Ascending  & 1976.20--1979.95, 1996.37--2000.28\\
  IV)  & $A>0$ Descending & 1968.87--1976.20, 1989.54--1996.37\\
  \hline
 \end{tabular}
\caption{Definition of Ascending and Descending periods}\label{tab::period}
\end{center}
\end{table}
\begin{figure}[htbp]
\centering
\includegraphics[width=0.9\textwidth]{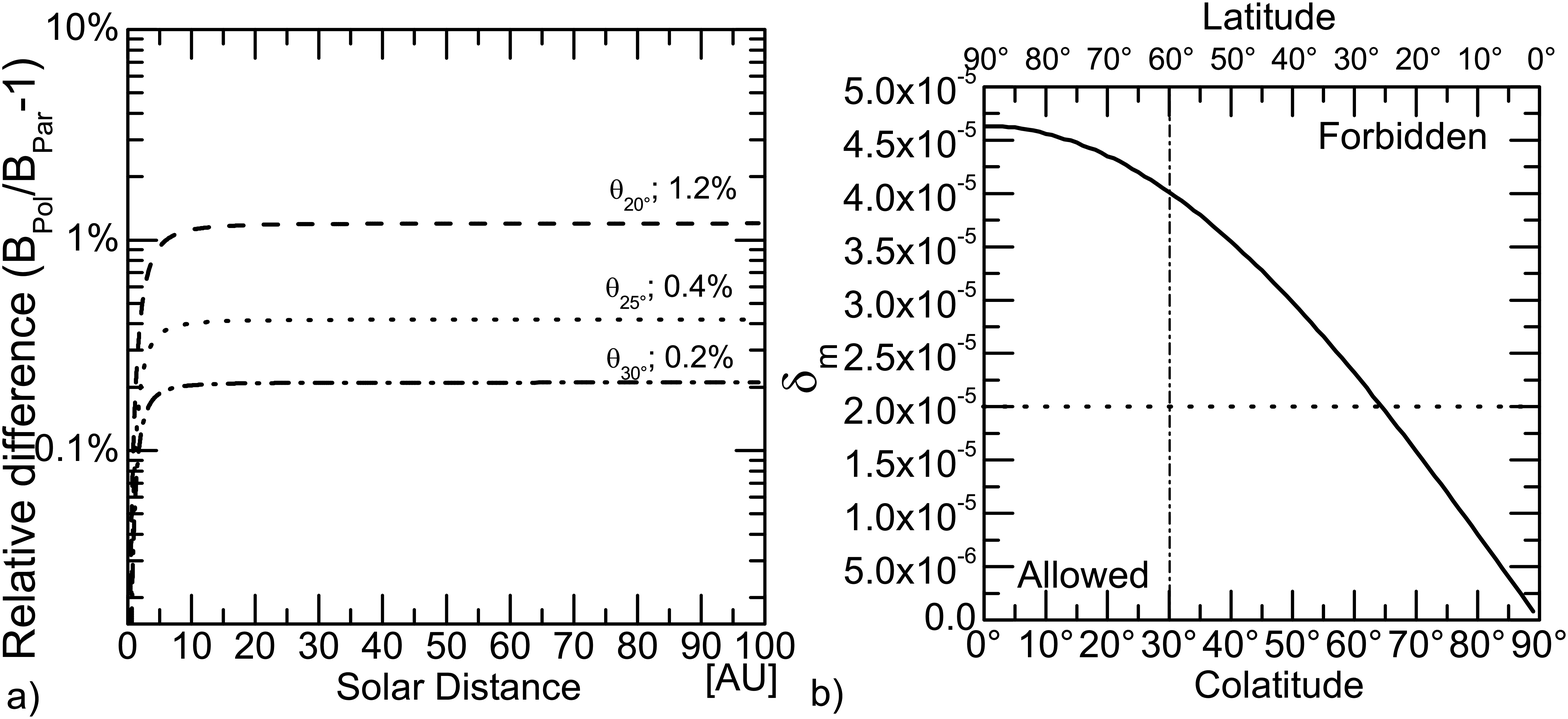}
\caption{(a) Maximum percentage difference between $B_{Pol}$ and $B_{Par}$ as a function of the solar distance inside
the colatitude regions $20\degree \leq \theta \leq 160\degree$ ($\theta_{20\degree}$),
$25\degree\leq \theta \leq 155\degree$ ($\theta_{25\degree}$) and
$30\degree\leq \theta \leq 150\degree$ ($\theta_{30\degree}$).
(b) Maximum value allowed of  $\delta_{m}$ as a function of colatitude: values in the ``allowed'' region guarantee
that stream lines originated from the polar magnetic field do not cross the equatorial plane.
Since Eq.~\eqref{IMFparpol} is symmetric
with respect to the solar equatorial plane, values greater than 90\degree~of colatitude lead to same results
of those presented.}
\label{BpolBpar}\label{deltam_max}
\end{figure}

In 1989~\cite{JokipiiKota89}, Jokipii and K\'ota have argued that the solar surface, where the
\textit{feet} of the field lines lie, is not a smooth surface, but a granular turbulent surface that keeps changing with
time, especially in the polar regions.
This turbulence may cause the \textit{footpoints} of the polar field lines to wander
randomly, creating transverse components in the field, thus causing temporal deviations from the smooth Parker
geometry. The net effect of this is a highly irregular and compressed field line. In other words, the magnitude
of the mean magnetic field at the poles is greater than in the case of the smooth magnetic field of a pure Parker
spiral. Jokipii and K\'ota~\cite{JokipiiKota89} have therefore suggested that the Parker spiral field may be generalized by the introduction
of a perturbation parameter [$\delta(\theta)$] which amplifies the field strength at large radial distances.
With this modification the magnitude of IMF, Eq.\eqref{eq::parkerBP_mag}, becomes \cite{JokipiiKota89}:
\begin{equation}\label{eq:JKMAg}
 B_{Pol} =\frac{A}{r^2}\sqrt{1+\Gamma^2+\left(\frac{r}{r_b}\right)^2\delta(\theta)^2}.
\end{equation}
The difference of the IMF obtained from Eq.~\eqref{eq::parkerBP_mag} and Eq.~\eqref{eq:JKMAg}
is less than $\sim$1\% for colatitudes $20\degree \leq  \theta\leq 160\degree$
(e.g, see Fig.~\ref{BpolBpar}a)
and increases for colatitudes approaching the polar regions
(e.g., see Figure 2 of~\cite{Bobik2011ApJ}).

In the present treatment, the heliosphere is divided into \textit{polar} regions and a \textit{equatorial} region where different description of IMF are applied.
In the \textit{equatorial} region the Parker's IMF, Eq.~\eqref{eq::parkerBP_mag}, is used, while in the \textit{polar} regions we used
a modified IMF that allows a magnitude as in Eq.~\eqref{eq:JKMAg}:
\begin{equation}\label{IMFparpol}
\left\{
\begin{array}{lll}
        \V B_{Pol} &= \dfrac{A}{r^{2}}\left[\V e_{r}+\dfrac{r}{r_b}\delta(\theta)\V e_{\theta}-\dfrac{\omega(r-r_b)\sin\theta}{V_{sw}}\V e_{\varphi}\right]\left[1-2H\left(\theta-\theta'\right)\right] & \textrm{Polar regions}\\
	\\\V B_{Par} &= \dfrac{A}{r^{2}}\left[\V e_{r}-\dfrac{\omega(r-r_b)\sin\theta}{V_{sw}}\V e_{\varphi}\right]\left[1-2H\left(\theta-\theta'\right)\right] \!\!& \textrm{Equatorial region},
\end{array}
\right.
\end{equation}
where equatorial regions are those with colatitude $X\degree \leq \theta \leq (180\degree-X\degree)$.
The symbol $\theta_{X\degree}$ indicates the corresponding polar regions.
\\ In order to have a divergence-free magnetic-field we require that the perturbation factor [$\delta(\theta)$] has to be:
\begin{equation}\label{deltatheta}
\delta(\theta)=\frac{\delta_{m}}{\left[1-2H\left(\theta-\theta'\right)\right]\sin\theta},
\end{equation}
where $\delta_{m}$ is the minimum perturbation factor of the field. The perturbation parameter is let to grow with decreasing
of the colatitude. However, in their original work, Jokipii and K\'ota~\cite{JokipiiKota89} estimated the value of the parameter
$\delta$ between $10^{-3}$ and $3\times10^{-3}$.

Since the polar field is only a perturbation of the Parker field, it is a reasonable assumption that
stream lines of the magnetic field do not cross the equatorial plane, thus, remaining completely contained in the solar hemisphere of injection.
This allows one to estimate an upper limit on the possible values of $\delta_{m}$ [see Fig.~\ref{deltam_max}(b)]. Currently, we use
$\delta_{m}=1\times10^{-5}$ by comparing simulations with observations at Earth orbit during Solar Cycle 23
(see Sect.~\ref{sect:DataComp}).

\begin{table}[tbp]
\begin{center}
 \begin{tabular}{cc}
\hline
  HelMod Energy bin & KET channel\\
\hline
  (0.35 -- 0.98)\,GeV   & (0.4 -- 1.0)\,GeV\\
  (0.76 -- 2.09)\,GeV   & (0.8 -- 2.0)\,GeV\\
  (2.09 -- 200)\,GeV    & $> 2$\,GeV\\
\hline
 \end{tabular}
\caption{Kinetic energy bins (in GeV) selected with HelMod Code and the corresponding proton energy channel
for KET instruments on board the Ulysses spacecraft~\cite{Heber1996}.}\label{tab:binHeber}
\end{center}
\end{table}
\begin{table}[tbhp]
\begin{center}
\begin{tabular}{cccccccccc}
\hline
  Energy range     & \multicolumn{3}{c}{$\theta_\textrm{lat,min}$ (in degrees)   } \\
                   & $\theta_{20\degree}$  & $\theta_{25\degree}$ & $\theta_{30\degree}$\\
\hline
  0.35 -- 0.98 GeV &  $-5^{+4}_{-5}$ & $-8^{+4}_{-4}$  & $-9^{+3}_{-3}$  \\
  0.76 -- 2.09 GeV & $-5^{+6}_{-7}$ & $-6^{+5}_{-6}$  & $-10^{+4}_{-4}$  \\
  2.09 -- 200 GeV  &$-3^{+7}_{-7}$ & $-7^{+7}_{-7}$& $-9^{+5}_{-6}$  \\
\hline
\end{tabular}
\caption{Latitudinal positions of minimum proton intensity ($\theta_\textrm{lat,min}$) (in degrees)
as a function of the kinetic energy,
 using three values for the extension of the polar regions. }\label{tab:shift}
\end{center}
\end{table}
\begin{table}[tbhp]
\begin{center}
\begin{tabular}{cccccccccc}
\hline
Energy range& \multicolumn{3}{c}{ $\Delta_\textrm{N-S}$ (\%)   } \\
                   & $\theta_{20\degree}$  & $\theta_{25\degree}$ & $\theta_{30\degree}$\\
\hline
  0.35 -- 0.98 GeV & $-8^{+6}_{-6}$ & $-11^{+5}_{-4}$& $-13^{+4}_{-4}$ \\
  0.76 -- 2.09 GeV & $-5^{+5}_{-6}$ & $-6^{+5}_{-5}$& $-9^{+3}_{-3}$\\
  2.09 -- 200 GeV  & $-1^{+3}_{-1}$ & $-3^{+3}_{-3}$& $-4^{+2}_{-2}$\\
\hline
\end{tabular}
\caption{Percentages of North-South asymmetry of proton intensities ($\Delta_\textrm{N-S}$) as a function of the kinetic energy,
 using three values for the extension of the polar regions. }\label{tab:Asym}
\end{center}
\end{table}

\begin{table}[tbhp]
\begin{center}
\begin{tabular}{cccccccccc}
\hline
Energy range& \multicolumn{3}{c}{$\Delta_\textrm{max}$ (\%)   } \\
                   & $\theta_{20\degree}$  & $\theta_{25\degree}$ & $\theta_{30\degree}$\\
\hline
  0.35 -- 0.98 GeV &  $-34^{+5}_{-5}$& $-35^{+4}_{-4}$ & $-36^{+4}_{-3}$\\
  0.76 -- 2.09 GeV &  $-22^{+6}_{-7}$& $-23^{+5}_{-4}$   & $-25^{+3}_{-3}$   \\
  2.09 -- 200 GeV  &  $-8^{+4}_{-3}$ & $-9^{+3}_{-3}$& $-10^{+2}_{-2}$\\
\hline
\end{tabular}
\caption{Differences in percentage between the maximum and minimum proton intensities ($\Delta_\textrm{max}$)
as a function of the kinetic energy,
 using three values for the extension of the polar regions. }\label{tab:deep}
\end{center}
\end{table}

\section{The Propagation Model}
Parker in 1965 \cite{parker1965} (see also Ref. \cite{Bobik2011ApJ}, Chapter 4 of~\cite{rancoita2011} and references therein) treated
the propagation of GCRs trough the
interplanetary space. He accounted for the so-called adiabatic energy losses, outward convection due to the SW and drift effects.
In the heliocentric system the Parker equation is then expressed (e.g. see~\cite{Bobik2011ApJ,Jokipii77}):
\begin{equation}\label{EQ::FPE_gen}
 \frac{\partial \mathrm{U}}{\partial t}= \frac{\partial}{\partial x_i} \left( K^S_{ij}\frac{\partial \mathrm{U} }{\partial x_j}\right)
- \frac{\partial}{\partial x_i} [ (V_{ \mathrm{sw},i}+v_{d,i})\mathrm{U}]
+\frac{1}{3}\frac{\partial V_{ \mathrm{sw},i} }{\partial x_i} \frac{\partial }{\partial T}\left(\alpha_{\mathrm{rel} }T\mathrm{U} \right)
\end{equation}
where $\mathrm{U}$ is the number density of particles per unit of particle kinetic energy $T$, at the time $t$. $V_{ \mathrm{sw},i}$
is the solar wind velocity along the axis $x_i$~\cite{Marsch2003}, $K^S_{ij}$ is the symmetric part of diffusion tensor~\cite{parker1965},
$v_{d}$ is the drift velocity that takes into account the
drift of the particles due to the large scale structure of the magnetic field~\cite{JokipiLev1977,Potgieter85,BurgerHatting1995} and, finally,
\[\alpha_{\mathrm{rel}}=\frac{T+2m_r c^2}{T+m_r c^2},\]
where $m_r$ is the rest mass of the GCR particle.~The last term of Eq.~\eqref{EQ::FPE_gen} accounts for adiabatic energy losses~\cite{parker1965,Jokipii1970}.
The number density $\mathrm{U}$ is related to the differential intensity $J$ as~(\cite{Bobik2011ApJ,FFM1968}, Chapter 4 of~\cite{rancoita2011} and references therein):
\begin{equation}
 J=\frac{vU }{4\pi},
\end{equation}
where $v$ is the speed of the GCR particle.
\begin{figure}[htbp]
\centering
\includegraphics[width=1\textwidth]{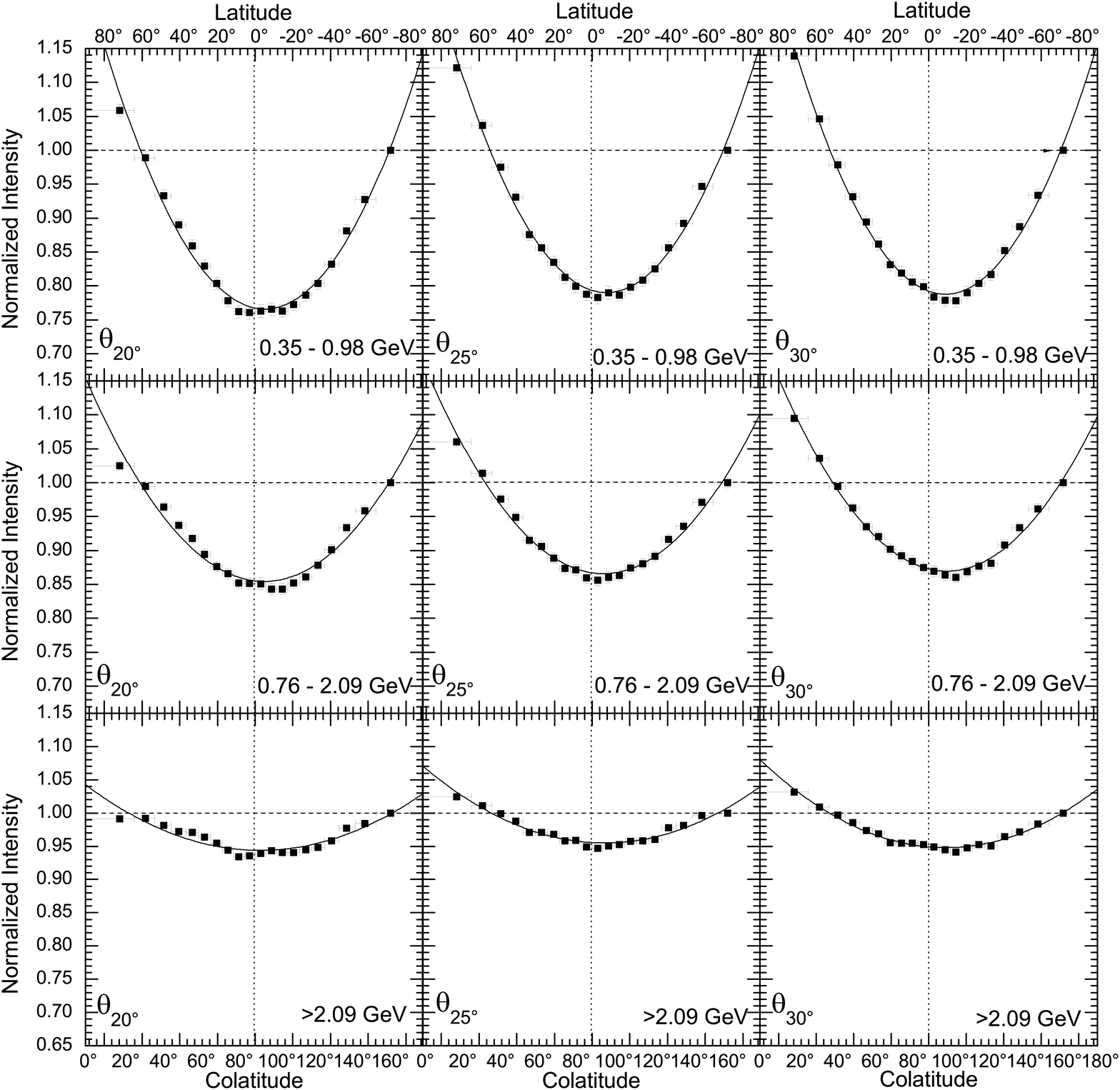}
\caption{Latitudinal relative intensity at $r=$\,1AU, obtained
at different solar colatitudes for protons in the energy range defined in the Table~\ref{tab:binHeber}
and using three definitions of \textit{polar regions}: $\theta_{20\degree}$, $\theta_{25\degree}$ and $\theta_{30\degree}$.  }
\label{GradRatio}
\end{figure}
\begin{figure}[htbp]
\centering
\includegraphics[width=0.7\textwidth]{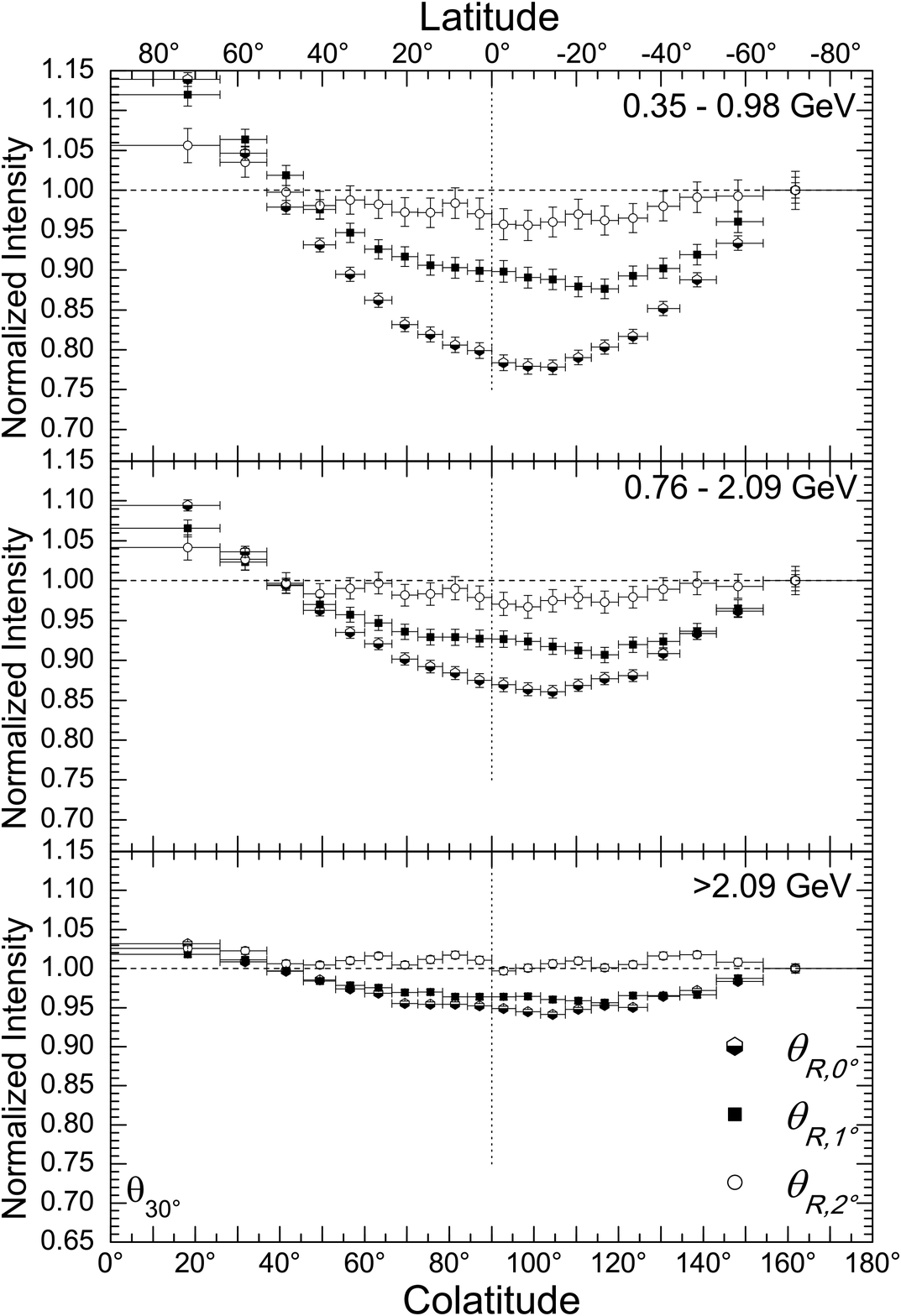}
\caption{For the
\textit{polar regions} $\theta<30\degree$ and $\theta>150\degree$, latitudinal relative intensity at $r=$1\,AU, accounting for GCR particles with
$0\degree<\theta<180\degree$ ($\theta_\textrm{R,0}$), $1\degree<\theta<179\degree$ ($\theta_\textrm{R,1}$) and
 $2\degree<\theta<178\degree$ ($\theta_\textrm{R,2}$), respectively, are shown as a function of the proton kinetic energy and solar colatitude. Results with
$10\degree<\theta<170\degree$ ($\theta_\textrm{R,10}$) are comparable with those obtained with $\theta_\textrm{R,2}$.}
\label{GradRatio_helio}
\end{figure}

Equation~\eqref{EQ::FPE_gen} was solved using the HelMod code (see the discussion in Ref.~\cite{Bobik2011ApJ}). This treatment
(i) follows that introduced in Refs. \cite{Yamada1998,GervasiEtAl1999,Zhang1999,alanko2007,PeiBurger2010,Strauss2011}
and (ii) determines the differential intensity of GCRs using
a set of approximated stochastic differential equations (SDEs) which provides a solution equivalent to that from Eq.~\eqref{EQ::FPE_gen}.
The equivalence between the Parker equation, that is a Fokker-Planck type equation, and the SDEs is demonstrated in \cite{Zhang1999, Gardiner1985}.
In the present work, we use a 2D (radius and colatitude) approximation for the particle transport. The model includes the effects of solar activity during the
propagation from the effective boundary of the heliosphere down to Earth's position.

The set of SDEs for the 2D approximation of Eq.~\eqref{EQ::FPE_gen} in heliocentric
spherical coordinates is
\begin{subequations}\label{eq::SDE_Hel2}
\begin{eqnarray}
\Delta r &=&  \frac{1}{r^2}\frac{\partial}{\partial r}(r^2 K^S_{rr})\Delta t - \frac{\partial}{\partial \mu(\theta)}\left[\frac{ K^S_{r\mu}\sqrt{1-\mu^2(\theta)}}{r}\right]\Delta t \nonumber \\
         &&   + ( V_{\rm{sw}}+v_{d,r})\Delta t+   \left(2K^S_{rr}\right)^{1/2} \omega_{r} \sqrt{\Delta t},\\
\Delta \mu(\theta)&= & -\frac{1}{r^2} \frac{\partial}{\partial r}\left[r K^S_{\mu r}\sqrt{1-\mu^2(\theta)}\right]\Delta t+ \frac{\partial}{\partial \mu(\theta)}\left[ K^S_{\mu \mu}\frac{1-\mu^2(\theta)}{r^2}\right]\Delta t \nonumber \\
       && -\frac{1}{r}v_{d,\mu}\sqrt{1-\mu^2(\theta)}\Delta t -  \frac{ 2 K^S_{r\mu}}{r}	 \left[\frac{1-\mu^2(\theta)}{2K^S_{rr}} \right]^{1/2} \omega_{r} \sqrt{\Delta t} \nonumber \\
      &&+ \frac{1}{r} \left\{ [1-\mu^2(\theta)]\frac{K^S_{\mu\mu}K^S_{rr}-(K^S_{r\mu})^2 }{0.5 K^S_{rr}}  \right\}^{1/2}  \omega_{\mu} \sqrt{\Delta t},\\
\Delta T &=&  -\frac{\alpha_{\rm rel} T}{3 r^2}\frac{\partial V_{\rm sw}r^2}{\partial r}   \Delta t, \label{eq::SDE_Hel2_deltaT}
\end{eqnarray}
\end{subequations}
where  $\mu(\theta)=\cos\theta$ and
$\omega_i$ is a random number following a Gaussian distribution with a mean of zero and a
standard deviation of one.
The procedure for determining the SDEs can be found in~\cite{Bobik2011ApJ}.

In a coordinate system with one axis parallel to the average magnetic field and the other two perpendicular to this
the symmetric part of the diffusion tensor $K^S_{ij}$ is  (see e.g.~\cite{jokipii1971}):
\begin{equation}\label{eq:Kmag}
 K^S_{ij}=\left [\begin{array}{ccc}
          K_{||} &   0     & 0 \\
           0      & K_{\perp,r} & 0  \\
         0           & 0          & K_{\perp,\theta}
        \end{array}\right ]
\end{equation}
with $ K_{||}$ the diffusion coefficient describing the diffusion parallel to the average magnetic field, and $K_{\perp,r}$ and
$K_{\perp,\theta}$ are the diffusion coefficients describing the diffusion perpendicular to the average magnetic field
in the radial and polar directions, respectively.
In this work $K_{||}$ is that one proposed by Strauss and collaborators in Ref.~\cite{Strauss2011} (see also~\cite{Palmer1982,PotgieterFerreira2002,droge2005}):
\begin{equation}\label{EQ::KparActual}
 K_{||}=\frac{\beta}{3} K_0 \frac{P}{1\text{GV}} \left(1+\frac{r}{\text{1 AU}}\right),
\end{equation}
where  $K_0$ is the diffusion parameter - described in Section 2.1 of~\cite{Bobik2011ApJ} -, which depends on solar activity and polarity, $\beta$ is the particle speed in unit of speed of light, $P=pc/|Z|e$ is the particle rigidity expressed in GV and, finally, $r$ is the
heliocentric distance from the Sun in AU.

In the current treatment, $K_{||}$ has a radial dependence proportional to $ r$, but no latitudinal dependence.
Mc Donald and collaborators (see Ref.~\cite{McDonald1997}) remarked that i) a spatial dependence of $K_{||}$ - like the one proposed here -
can affect the latitudinal gradients at high latitude and ii) it
is consistent with that originally suggested in Ref.~\cite{JokipiiKota89}.
Furthermore, the perpendicular diffusion coefficient is taken to be proportional to $K_{||}$ with a ratio
$K_{\perp,i}/K_{||}=0.13$ for both $r$ and $\theta$ $i$-coordinates. The latter value is discussed in Sect.~\ref{sect:DataComp}.

In addition, the
practical relationship between $K_0$ and monthly Smoothed Sunspot Numbers (SSN) \cite{SSN} values
 - discussed in Section 2.1 of~\cite{Bobik2011ApJ} -
is currently updated using the most recent data from Ref. \cite{usoskin2011}.
As in Ref.~\cite{Bobik2011ApJ}, the $K_0$ data are subdivided into four sets, i.e.,
ascending and descending phases for both negative and positive
solar magnetic-field polarities (Table~\ref{tab::period}).
It has to be remarked that after each maximum the sign of the magnetic field (i.e., the $A$ parameter in Eq.~\eqref{IMFparpol})
is reversed.
The updated practical relationships between $K_0$ in $\textrm{AU}^2\textrm{GV}^{-1}\,\textrm{s}^{-1}$
and SSN values for $1.4\leq$SSN$\leq 165$ for the four periods (from I up to IV, listed in Table~\ref{tab::period}) are
\begin{subequations}\label{eq:practicalK0}
\begin{eqnarray}
\textrm{I)} & K_0=0.000297\!-\!2.9\!\cdot\! 10^{-6}\text{SSN}\!+\!8.1\!\cdot\! 10^{-9}\text{SSN}^2\!+\!1.46\!\cdot\! 10^{-10}\text{SSN}^3\!-\!8.4\!\cdot\!
 10^{-13}\text{SSN}^4  \\
\textrm{II)}&  K_0=\left \{ \begin{array}{ll}
                                                 0.000304846-5.8\cdot10^{-6}\text{SSN} & \text{if SSN}<=20\\
                                                \frac{0.00195}{\text{SSN}}-2.3\cdot 10^{-10}\text{SSN}^2+9.1\cdot 10^{-5} & \text{if SSN}>20
                                                \end{array}\right.  \\
\textrm{III)} &  K_0=0.0002391-8.453\cdot 10^{-7}\text{SSN} \\
\textrm{IV)} & K_0=0.000247-1.175\cdot 10^{-6}\text{SSN}.
\end{eqnarray}
\end{subequations}
The rms (root mean square) values of the percentage difference between values obtained with Eqs.~\eqref{eq:practicalK0} from those determined using the procedure discussed in Section 2.1 of~\cite{Bobik2011ApJ} applied to the data from Ref. \cite{usoskin2011} were found to be
6.0\%,  10.1\%, 7.0\% and 13.2\% for the period I (ascending phase with $A<0$), II
(descending phase with $A<0$), III (ascending phase with $A>0$) and IV (descending phase with $A>0$), respectively.

\begin{table}[htbp]
\begin{center}
\begin{tabular}{cc|rrrr}
\hline \hline
\multicolumn{1}{l}{$\delta_m$ $(\times 10^{-5})$} & \multicolumn{1}{l}{$\rho_k$} & \multicolumn{1}{l}{ ``R'' model   } & \multicolumn{1}{l}{ ``L'' model  } & \multicolumn{1}{l}{ No Drift }\\
\hline
0.0 & 0.10 & 14.1 & 11.0 & 33.4\\
1.0 & 0.10 & 11.7 & 8.7 & 33.2\\
2.0 & 0.10 & 11.6 & 8.3 & 33.7\\
3.0 & 0.10 & 11.6 & 8.3 & 33.7\\
1.0 & 0.11 & 6.4 & 9.0 & 27.7\\
2.0 & 0.11 & 7.8 & 9.0 & 28.3\\
3.0 & 0.11 & 7.5 & 8.8 & 29.2\\
1.0 & 0.12 & 6.3 & 7.1 & 23.5\\
2.0 & 0.12 & 6.3 & 7.3 & 24.7\\
3.0 & 0.12 & 7.1 & 6.9 & 24.4\\
1.0 & 0.13 & 6.3 & 6.4 & 20.1\\
2.0 & 0.13 & 6.6 & 7.6 & 20.4\\
3.0 & 0.13 & 6.7 & 7.7 & 20.5\\
1.0 & 0.14 & 7.3 & 7.0 & 15.9\\
2.0 & 0.14 & 7.3 & 7.2 & 16.4\\
3.0 & 0.14 & 7.2 & 6.5 & 16.8\\
\hline \hline
\end{tabular}
\caption{Average values (last three columns) of $\eta_\mathrm{rms}$ (in percentage, \%) as a function of $\delta_m$ $(\times 10^{-5})$ and $\rho_k$, for BESS--1997, AMS--1998, PAMELA-2006/08, obtained from Eq.~\eqref{eq::etaRms_dat}
without enhancement of the diffusion tensor along the polar direction ($\rho_E=1$), using ``R'' and  ``L'' models
for the tilt angle and No Drift approximation. The differential intensities were calculated accounting for particles
inside the heliospheric regions for which solar latitudes are lower than
$|5. 7\degree|$. }
\label{tab:averLowNoEn}
\end{center}
\end{table}
\begin{figure}[tbp]
\begin{center}
 \includegraphics[width=0.6\textwidth]{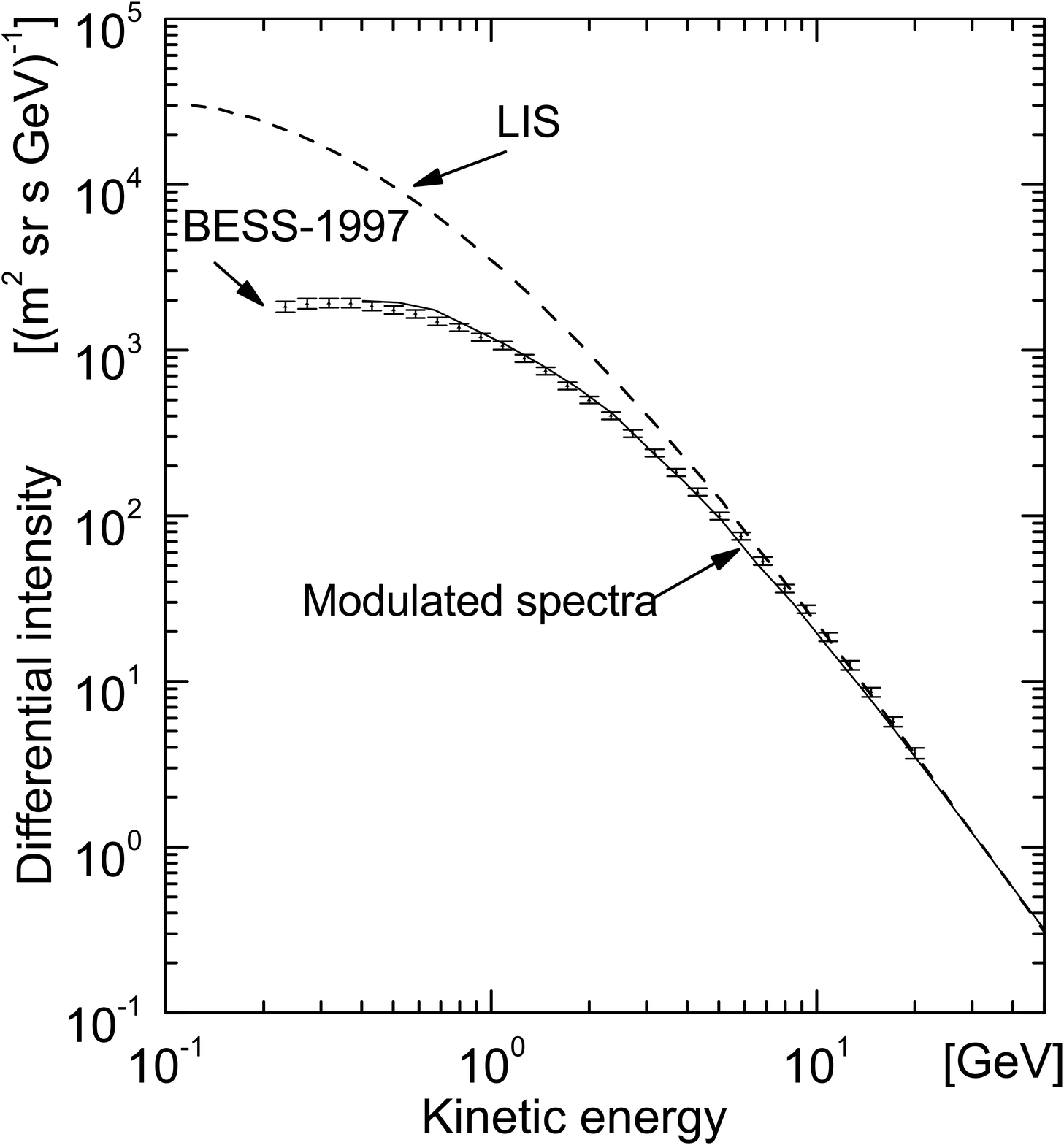}
 \caption{Proton differential intensity determined with the HelMod code (continuous
line) compared to the experimental data of BESS--1997; the dashed line is the
LIS (see the text).
}
 \label{fig:best_Bess97}
\end{center}
\end{figure}
\begin{figure}[tbp]
\begin{center}
 \includegraphics[width=0.6\textwidth]{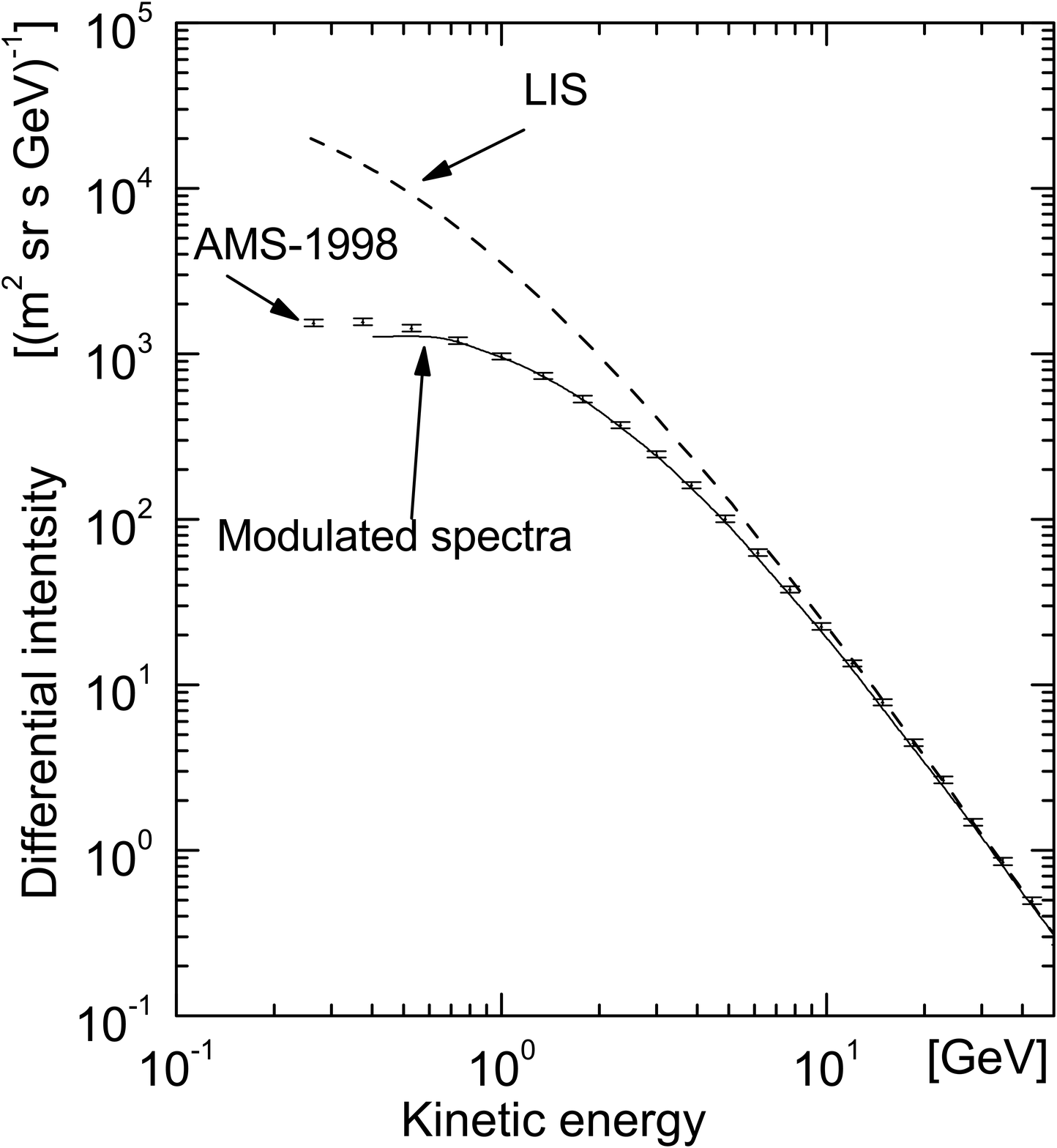}
 \caption{Proton differential intensity determined with the HelMod code (continuous
line) compared to the experimental data of AMS-1998; the dashed line is the
LIS (see the text).
}
 \label{fig:best_ams98}
\end{center}
\end{figure}
\begin{figure}[tbp]
\begin{center}
 \includegraphics[width=0.6\textwidth]{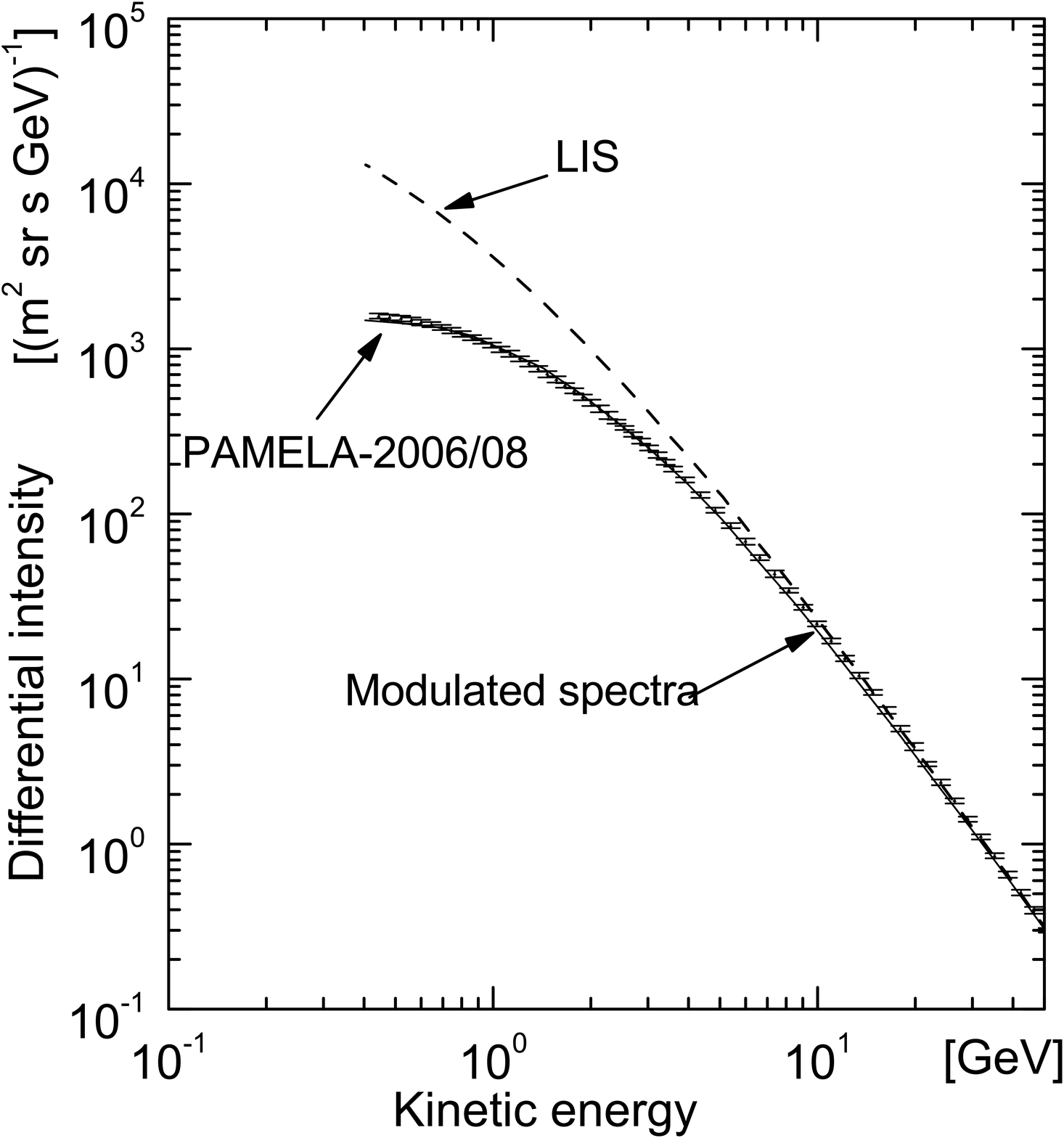}
 \caption{Proton differential intensity determined with the HelMod code (continuous
line) compared to the experimental data of PAMELA 2006/08; the dashed line is the
LIS (see the text).
}
 \label{fig:best_PAMELA}
\end{center}
\end{figure}

\section{The Magnetic Field in the Polar Regions}
Section~\ref{IMF} describes an IMF following the Parker Field
with a small region around the poles in which
such a field is modified.
As already mentioned (see e.g.\cite{JokipiiKota89,BaloghEtAl1995,Moraal1990,SmithBieber1991,Fisk1996,HitgeBurger2010}), the correction is needed to better reproduce the
complexity of the magnetic field in those regions.

Moreover, Ulysses spacecraft (see e.g.~\cite{Sandersonetal1995,Marsden2001,BaloghetAl2001}) explored
the heliosphere outside the ecliptic plane up to $\pm 80\degree$ of solar latitude at a solar distance
from $\sim 1$ up to $\sim 5$\,AU. Using these observations,
the presence of latitudinal gradient in the proton intensity could be determined
(e.g., see Figure 2 of Ref.~\cite{Simpson1996a} and Figure 5 of Ref. \cite{Heber1996}).
The data collected during the \textit{latitudinal fast scan}
(from September 1994 up to August 1995) show (a) a nearly symmetric latitudinal
gradient with the minimum near ecliptic plane, (b) a southward shift of the minimum
and (c) an intensity in the North polar region at $80\degree$ exceeding the
South polar intensity.
In Ref. \cite{Simpson1996a} a latitudinal gradient of $\sim 0.3\%$/degree~
for proton with kinetic energy $>0.1$ GeV was estimated. While in Ref. \cite{Heber1996} the analysis to higher energy
was extended estimating a gradient of $\sim 0.22\%$/degree
for proton with kinetic energy $>2$ GeV.
The minimum in the charged particle intensity
separating the two hemispheres of the heliosphere occurs
$\sim 10\degree$ South of the heliographic equator~\cite{Simpson1996a}. In addition,
an independent analysis that takes into account the latitudinal motion of the Earth and IMP8 confirms
a significant ($\sim 8\degree \pm 2\degree$) southward offset of the intensity
minimum~\cite{Simpson1996a} for  $T> 100$ MeV proton. Furthermore in Ref. \cite{Heber1996}, a southward offset
of about $\approx 7\degree$ is evaluated; this offset of the intensity minimum results to be independent of
the particle energy up to 2 GeV.
Finally, in Ref. \cite{Simpson1996a}, the intensity in the North polar region at $80\degree$ is observed to exceed the
South polar intensity of $\sim 6\%$ for protons with $T> 100$ MeV.
\begin{table}[htbp]
\begin{center}
\begin{tabular}{cc|rrrr}
\hline \hline
\multicolumn{1}{l}{$\delta_m$ $(\times 10^{-5})$} & \multicolumn{1}{l}{$\rho_k$} & \multicolumn{1}{l}{ ``R'' model   } & \multicolumn{1}{l}{ ``L'' model  } & \multicolumn{1}{l}{ No Drift }\\
\hline
0.0 & 0.10 & 11.2& 10.8 & 15.4 \\
1.0 & 0.10 & 11.0& 10.1 & 15.8 \\
2.0 & 0.10 & 9.6 & 10.0 & 16.7 \\
3.0 & 0.10 & 9.6 & 10.0 & 16.7 \\
1.0 & 0.11 & 13.4& 13.1 & 16.0 \\
2.0 & 0.11 & 12.7& 12.9 & 15.4 \\
3.0 & 0.11 & 12.7& 12.5 & 16.2 \\
1.0 & 0.12 & 18.7& 17.7 & 13.4 \\
2.0 & 0.12 & 18.3& 16.9 & 12.8 \\
3.0 & 0.12 & 18.1& 17.3 & 12.8 \\
1.0 & 0.13 & 23.3& 23.5 & 14.3 \\
2.0 & 0.13 & 25.0& 24.7 & 13.3 \\
3.0 & 0.13 & 24.3& 24.2 & 13.1 \\
1.0 & 0.14 & 32.3& 30.7 & 18.0 \\
2.0 & 0.14 & 32.8& 30.8 & 17.1 \\
3.0 & 0.14 & 31.5& 30.7 & 17.9 \\
\hline \hline
\end{tabular}
\caption{Average values (last three columns) of $\eta_\mathrm{rms}$ (in percentage, \%) as a function of $\delta_m$ $(\times 10^{-5})$ and $\rho_k$, for BESS--1999, BESS--2000, BESS--2002, obtained from Eq.~\eqref{eq::etaRms_dat}
 without enhancement of the diffusion tensor along the polar direction ($\rho_E=1$), using ``R'' and  ``L'' models
for the tilt angle and No Drift approximation. The differential intensities were calculated accounting for particles
inside the heliospheric regions for which solar latitudes are lower than
$|5.7\degree|$. }
\label{tab:averHighNoEn}
\end{center}
\end{table}
\begin{figure}[tbp]
\begin{center}
 \includegraphics[width=0.6\textwidth]{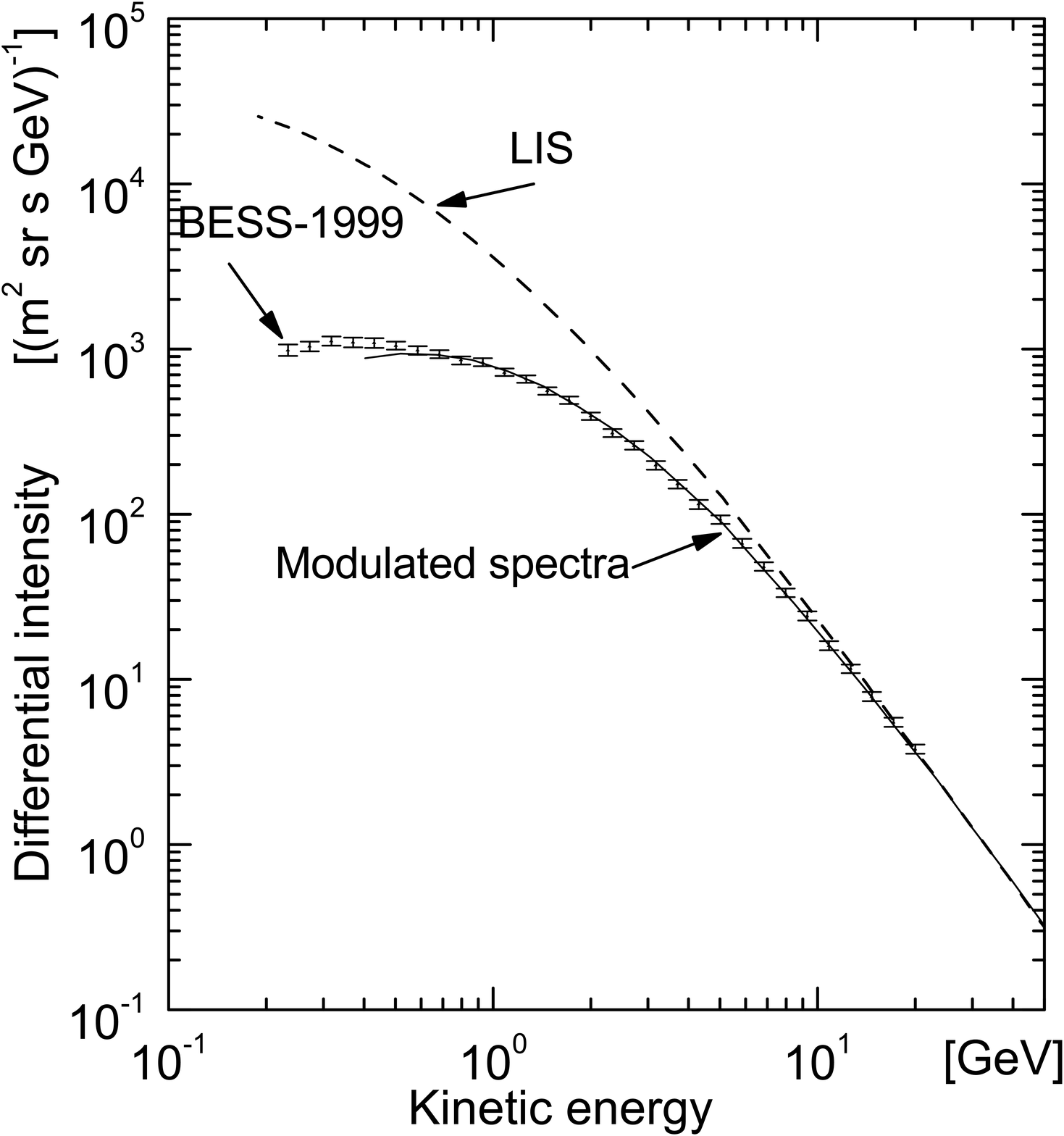}
 \caption{Proton differential intensity determined with the HelMod code (continuous
line) compared to the experimental data of BESS--1999; the dashed line is the
LIS (see the text).
}
 \label{fig:best_BESS1999}
\end{center}
\end{figure}
\begin{figure}[tbp]
\begin{center}
 \includegraphics[width=0.6\textwidth]{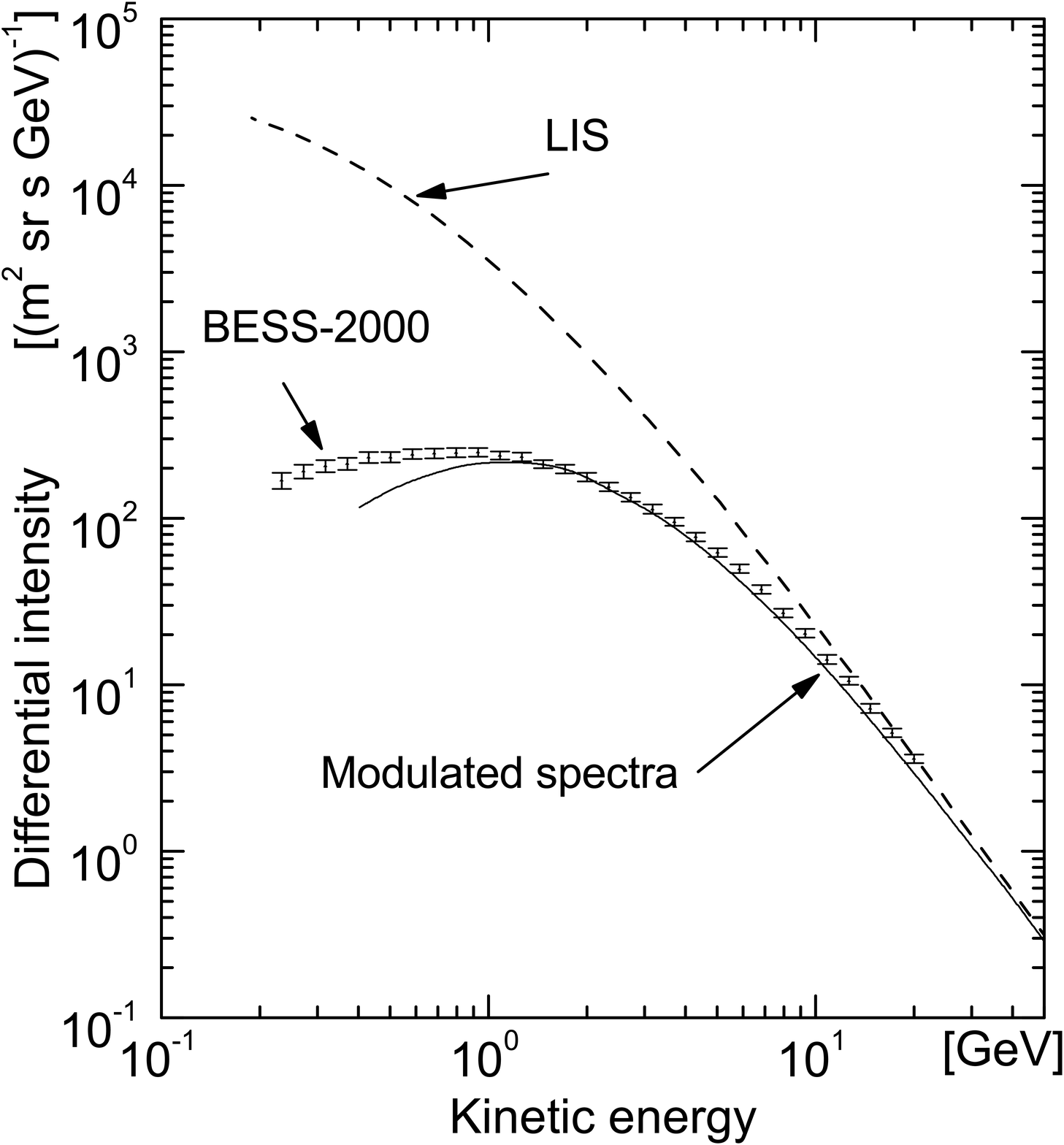}
 \caption{Proton differential intensity determined with the HelMod code (continuous
line) compared to the experimental data of BESS--2000; the dashed line is the
LIS (see the text).
}
 \label{fig:best_BESS2000}
\end{center}
\end{figure}
\begin{figure}[tbp]
\begin{center}
 \includegraphics[width=0.6\textwidth]{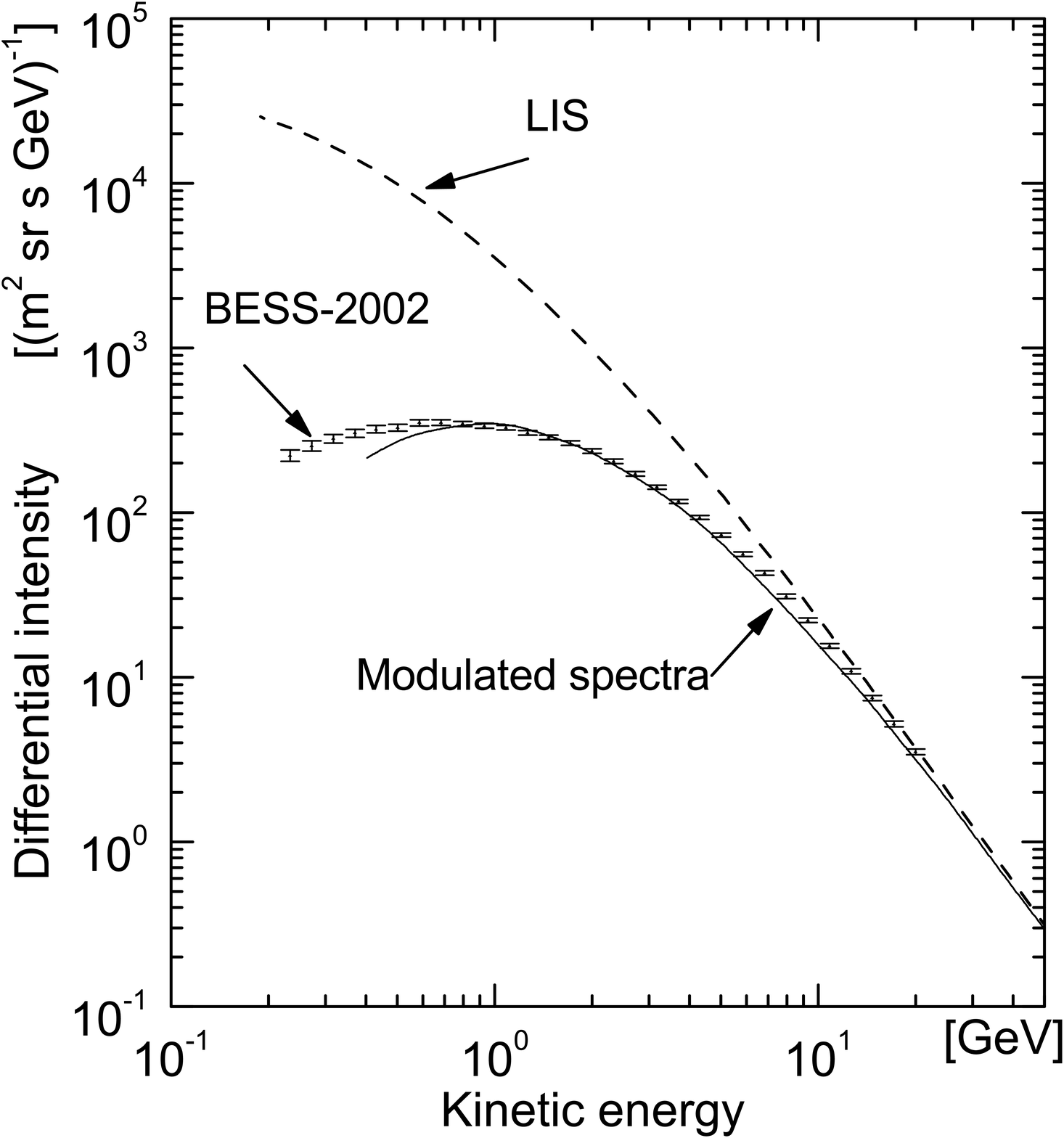}
 \caption{Proton differential intensity determined with the HelMod code (continuous
line) compared to the experimental data of BESS--2002; the dashed line is the
LIS (see the text).
}
 \label{fig:best_BESS2002}
\end{center}
\end{figure}

Using the present HelMod code (version 2.0), we could investigate i) the latitudinal gradient of GCR intensities resulting from solar modulation and ii)
how the magnetic-field structure of the polar regions, as defined in Sect.~\ref{IMF}, is able to
influence the GCR spectra on the ecliptic plane.
As previously defined, we denote with $\theta_{X\degree}$ a polar region
of amplitude $X\degree$ from polar axis, i.e., $\theta<X\degree$ and $\theta>180\degree-X\degree$.
Three regions with $X\degree = 20\degree,\,25\degree$ and 30\degree\,
were investigated.
Outside any of these regions, the ratio between $B_{Pol}$ and $B_{Par}$ in Eq.~\eqref{IMFparpol}
is less than $\sim$1\% (see Fig.~\ref{deltam_max}) and, thus, it ensures a smooth transition between polar and equatorial regions.
%
For the purpose or this study we consider an energy binning closer to those presented in Ref. \cite{Heber1996}.
The KET instrument~\cite{Heber1997}
collects proton data in three ``channels'' one with energies ranging from 0.038\,GeV up to 2.0\,GeV and two
for particle with kinetic energy $T>0.1$ GeV and $T>2$ GeV, respectively. A successive re-analysis
of the collected data allowed the authors to subdivide the 0.25--2 GeV ``channel'' in three ``sub-channels'' of intermediate energies.
Since the Present Model is optimized - as discussed in Ref.~\cite{Bobik2011ApJ} - for particles with rigidity greater than 1\,GV (i.e., $\approx 0.444$\,GeV),
the present results are compared only with the corresponding ``channel'' or ``sub-channels'' suited for the corresponding energy range (see Table \ref{tab:binHeber}).

At 1\,AU and as a function of the solar colatitude, the GCR intensities for protons are shown in Fig.~\ref{GradRatio}. For a comparison with Ulysses observations,
the modulated intensities of protons - resulting from HelMod code - were investigated from $80\degree$ (North) and down to $-80\degree$ (South). They were
obtained using the HelMod code and selected using the energy bins reported in table \ref{tab:binHeber}. In Fig.~\ref{GradRatio}, the latitudinal intensity distribution is normalized to the corresponding South Pole intensity.
The quoted errors include statistical and systematic errors.
The distributions were interpolated using a parabolic function expressed as:
\begin{equation}\label{eq:fit}
I(\theta_\textrm{lat})=a+c(\theta_\textrm{lat} + d)^2,
\end{equation}
where $I(\theta_\textrm{lat})$ is the normalized intensity, $\theta_\textrm{lat}$
is the latitudinal angle\footnote{The latitudinal angle is $\theta_\textrm{lat}=90\degree-\theta$.}
and $a,b$ and $c$ are parameters determined from the fitting procedure. The so obtained fitted curves are shown as continuous lines in Fig.~\ref{GradRatio}. Furthermore, the latitudinal positions of minimum intensity ($\theta_\textrm{lat,min}$), percentages of North-South asymmetry of intensities ($\Delta_\textrm{N-S}$) and differences in percentage between the maximum and minimum intensities ($\Delta_\textrm{max}$) were also determined from the fitting procedure and are listed in Table~\ref{tab:shift}, Table~\ref{tab:Asym} and Table~\ref{tab:deep}, respectively.
The quoted errors - following a procedure discussed in Ref.~\cite{Bobik2011ApJ} - are derived varying the fitted parameters in order to obtain
a value of the parameter $\eta_\mathrm{rms,lat}$ two times larger than the one
resulting from the best fit ($I_\mathrm{best}$).
$\eta_\mathrm{rms,lat}$ is defined as:
\begin{equation}\label{eq::etaRms_lat}
 \eta_\mathrm{rms.lat}=\sqrt{\frac{\sum_i (\eta_{i,lat}/\sigma_{\eta,i,lat})^2}{\sum_i 1/\sigma^2_{\eta,i,lat}}}
\end{equation}
with
\begin{equation}\label{eq::etai_lat}
 \eta_{i,lat}=\frac{I(\theta_i)-I_\mathrm{best}(\theta_i)}{I_\mathrm{best}(\theta_i)},
\end{equation}
where $\theta_i$ is the central value  of the $i$th latitudinal bin of
the differential intensity distribution and $\sigma_{\eta,i,lat}$ are the errors
including the experimental and Monte Carlo uncertainties.
For $\theta_{30\degree}$, i.e., assuming a modified polar magnetic-field for $\theta < 30\degree$ and $\theta > 150\degree$, we found a general agreement with Ulysses observations. The position of $\theta_\textrm{lat,min}$
is compatible within the errors with one observed in Ref.~\cite{Heber1996}, as well as the values
of $\Delta_\textrm{N-S}$ and $\Delta_\textrm{max}$.

The HelMod code allows one to investigate the relevance
of the treatment of the polar region magnetic-field with respect to the resulting modulation of GCR.
Thus, the latitudinal normalized intensities were obtained
excluding a few (small) regions nearby the poles.
This was determined from reducing the latitudinal spatial phase-space
admissible for \emph{pseudo-particles} (see Ref.~\cite{Bobik2011ApJ}) - i.e., the latitudinal extension of GCR particles taken into account -
to $1\degree<\theta<179\degree$ ($\theta_\textrm{R,1}$),
$2\degree<\theta<178\degree$ ($\theta_\textrm{R,2}$) and
$10\degree<\theta<170\degree$ ($\theta_\textrm{R,10}$). The so obtained latitudinal gradients are compared with the full latitudinal extension, $\theta_\textrm{R,0}$ ($0\degree<\theta<180\degree$), in Fig.~\ref{GradRatio_helio}.
By an inspection of Fig.~\ref{GradRatio_helio}, one may lead to the conclusion
that the GCR diffusion nearby the polar axis has a large impact on the latitudinal gradients
in the inner heliosphere. As a consequence, the IMF description in the polar regions is relevant in order to reproduce
the observed modulated GCR spectra.

\section{Comparison with Observations During Solar Cycle 23}\label{sect:DataComp}
The agreement of HelMod simulated spectra with observations
during solar cycle 23 is investigated via
quantitative comparisons using
Eqs. \eqref{eq::etaRms_dat} and \eqref{eq::etai_dat}. However,
since the structure of the heliosphere is different in high and low solar activity
the two periods are separately analyzed.

The HelMod Code~\cite{Bobik2011ApJ} (version 2.0) allowed us to investigate how the modulated (simulated)
differential intensities are affected by the (1) particle drift effect,
(2) polar enhancement of the diffusion tensor
along the polar direction ($K_{\perp,\theta}$, e.g., see Ref.~\cite{potgieter2000}), and, finally,
(3) values of the tilt angle ($\alpha_t$) calculated following the approach
of the ``R'' and ``L'' models \cite{Hoeksema1995}.
This analysis also allow us to estimate the values of IMF parameters that better describe
the modulation along the entire solar cycle. The effects related to particle drift were investigated
via the suppression of the drift velocity (\textit{No Drift} approximation), this accounts
for the hypothesis that magnetic drift convection is almost
completely suppressed during solar maxima.
The differential intensities were calculated
for $K_{\perp,\mu}= \rho_{E} K_{\perp,r} $
with values of $\rho_E$ of 1, 8 and 10, i.e., no  enhancement, that suggested in Ref. \cite{Ferreira2004}
and that suggested
in Ref. \cite{potgieter2000,Bobik2011ApJ} (and reference there in), respectively.
Furthermore, the modulated proton spectra
were derived from a LIS  whose
normalization constant depends on the experimental set
of data and were already discussed in Ref.~\cite{Bobik2011ApJ}).
In addition, the differential intensities were calculated accounting for particles
inside heliospheric regions where solar latitudes are lower than
$|5. 7\degree|$.

During the period of high solar activity for the solar cycle
23, the BESS collaboration took data in the years 1999, 2000,
and 2002 (see sets of data in Ref.~\cite{bess_prot}).
For period not dominated by high solar activity in solar
cycle 23, BESS, AMS and PAMELA collaborations took data, i.e., BESS--1997 \cite{bess_prot}, AMS--1998
\cite{AMS01_prot}, 
and PAMELA--2006/08 \cite{Pamela_Prot}.

Following the procedure described in Ref. \cite{Bobik2011ApJ}, the observation data
were compared with those obtained from HelMod
code using the error-weighted root mean square ($\eta_\mathrm{rms}$) of the
relative difference ($\eta$) between experimental data ($f_\mathrm{exp}$) and those
resulting from simulated differential intensities ($f_\mathrm{sim}$). For each
set of experimental data and with the approximations and/or
models described above, we determined the quantity
\begin{equation}\label{eq::etaRms_dat}
 \eta_\mathrm{rms}=\sqrt{\frac{\sum_i (\eta_i/\sigma_{\eta,i})^2}{\sum_i 1/\sigma^2_{\eta,i}}}
\end{equation}
with
\begin{equation}\label{eq::etai_dat}
 \eta_i=\frac{f_\mathrm{sim}(T_i)-f_\mathrm{exp}(T_i)}{f_\mathrm{ref}(T_i)},
\end{equation}
where $T_i$ is the average energy of the $i$th energy bin of
the differential intensity distribution and $\sigma_{\eta,i}$ are the errors
including the experimental and Monte Carlo uncertainties; the
latter account for the Poisson error of each energy bin.
The simulated differential intensities are interpolated with a cubic
spline function.
The modulation results are studied varying the parameters $\delta_m$ -
from 0, i.e., the non-modified Parker IMF, up to $3\times 10^{-5}$, see Fig.~\ref{deltam_max}(b) -,
$K_{\perp,r} / K_{||}=\rho_k$ (from 0.10 up to 0.14) and
$K_{\perp,\theta} / K_{\perp,r}=\rho_E$ (from 1 up to 10)
seeking a set of parameters set that minimize $\eta_\mathrm{rms}$.
In Table~\ref{tab:averLowNoEn},
the average values of $\eta_\mathrm{rms}$ (in percentage, \%) for low solar activity periods are listed.
They were obtained in the energy range~\footnote{Above 30\,GeV, the differential intensity is marginally (if at all) affected by
modulation.} from
444 MeV up to 30 GeV using the ``L'' and ``R'' models
for the tilt angle $\alpha_t$ and for the No Drift approximation
and without any enhancement of the diffusion tensor along the
polar direction ($K_{\perp,\mu}$).
The results derived with the enhancement of the diffusion tensor along the
polar direction indicate
that for $\rho_E=$ 8 and 10 one obtains a value of $\eta_\mathrm{rms}$ that is from 1.5 up
to 3 times larger with respect the case without enhancement\footnote{For a comparison, the \textit{scalar} approximation presented in Ref.~\cite{Bobik2011ApJ}, i.e.,
assuming that the diffusion propagation
is independent of magnetic structure,
leads to and average  $\eta_\mathrm{rms}$ of $\sim 15\%$.}.
From inspection of Table~\ref{tab:averLowNoEn}, one can remark that
the drift mechanism leads to a better agreement with
experimental data. Furthermore the, ``R'' and ``L'' models for tilt angles are comparable
within the precision of the method (discussed in Ref. \cite{Bobik2011ApJ}).
The minimum difference with respect to the experimental data occurs
when $\rho_k= 0.11 - 0.13$ and $\delta_m=1.0\times 10^{-5}$ for both ``R'' and ``L'' models,
with the ``L'' model slightly preferred to ``R''.

In Figs.~\ref{fig:best_Bess97}--\ref{fig:best_PAMELA}, the differential
intensities determined with the HelMod code are shown and
compared to the experimental data of BESS--1997, AMS--1998,
and PAMELA--2006/08, respectively; in these
figures, the dashed lines are the LIS as discussed in Ref.~\cite{Bobik2011ApJ}.
These modulated intensities are the ones calculated for a heliospheric
region where solar latitudes are lower than $|5. 7\degree|$, using
$\rho_k=0.13$, $\delta_m=1.0\times 10^{-5}$ and $\rho_E= 1$ with the ``L'' model.
Finally, one can remark that the present code, combining
diffusion and drift mechanisms, is also suited to describe the
modulation effect in periods when the solar activity is no longer
at the maximum.

In Table~\ref{tab:averHighNoEn} we present
the averages $\eta_\mathrm{rms}$ (in percentage, \%) during the periods dominated by high solar activity.
The simulated differential intensities were obtained for a heliospheric region where solar
latitudes are lower than $|5.7\degree|$ without any  enhancement of the diffusion tensor along the
polar direction ($K_{\perp,\mu}$).
The simulations with $\rho_E=$ 8 and 10 lead to $\eta_\mathrm{rms}$ comparable with those presented
in Table~\ref{tab:averHighNoEn}. However, using $\rho_E=1$ provides a better agreement
to experimental data at lower energy.
From inspection of Table~\ref{tab:averHighNoEn},
one can note that ``R'' and ``L'' models for tilt angles yield comparable results
within the precision of the method.
Furthermore the minimum difference with the experimental data occurs
when $\rho_k= 0.10$ and $\delta_m=(2.0$--$3.0) \times 10^{-5}$ with the``L'' model slightly preferred to ``R''. The HelMod parameter configuration, which
minimizes the difference to the experimental data, are
reported in Table~\ref{tab:averHighNoEn_tabBest}: one may remark that the No Drift approximation is (almost)
comparable to a drift treatment for both BESS--2000 and BESS--2002 with data collected during and after the
maximum of the solar activity. Apparently, the drift treatment is needed
in order to describe BESS--1999 with data taken during a period approaching the solar maximum.
\begin{table}[htbp]
\begin{center}
\begin{tabular}{rrrrr}
 \hline \hline
Observations & ``R'' model    & ``L'' model  & No Drift \\
\hline
BESS--1999   &9.3&	10.6&	25.7\\
BESS--2000   &12.5&	12.6&	16.7\\
BESS--2002   &6.9&	6.7&	7.7\\
\hline
\end{tabular}
\caption{Average $\eta_\mathrm{rms}$ (in percentage, \%), for BESS--1999, BESS--2000, BESS--2002, obtained from Eq.~\eqref{eq::etaRms_dat}
without enhancement of the diffusion tensor along the polar direction ($\rho_E=1$), $\delta_m=2.0\times 10^{-5}$, $\rho_k=0.10$
and using ``R'' and  ``L'' models
for the tilt angle and No Drift approximation. The differential intensities were calculated accounting for particles
inside the heliospheric regions for which solar latitudes are lower than
$|5.7\degree|$. }
\label{tab:averHighNoEn_tabBest}
\end{center}
\end{table}

In Figs.~\eqref{fig:best_BESS1999}--\eqref{fig:best_BESS2002},
the differential intensities
determined with the HelMod code are shown and compared
with the experimental data of BESS--1999, BESS--2000, and
BESS--2002, respectively; in these
figures, the dashed lines are the LIS as discussed in Ref.~\cite{Bobik2011ApJ}. These modulated intensities are the
ones calculated for a heliospheric region where solar latitudes
are lower than $|5.7\degree|$, using $K_{\perp,\mu} = K_{\perp,r}$ independently of the
latitude and including particle drift effects with the values of the
tilt angle from the ``L'' model.
Finally, it is concluded that the present code combining diffusion
and drift mechanisms is suited to describe the modulation
effect in periods with high solar activity \cite{Bobik2011ApJ,Ferreira2004,Ndiitwani2005}.

\section{Conclusion}
In this work an IMF, which combines the Parker Field and its polar modification, is presented.
In the polar regions, the Parker IMF was modified with an additional latitudinal components according to those proposed by
Jokipii and K\'ota in Ref.~\cite{JokipiiKota89}. We found the maximum perturbed value with this component yielding, as a physical result, streaming lines completely confined in the solar hemisphere of injection.

The proposed IMF is, then, used within the HelMod Monte Carlo code to
determine the effects on the differential intensity of protons at 1\,AU as a function of the extension
of \textit{polar region}, in which the modified magnetic-field is employed.
We found that a \textit{polar region} contained within 30\degree~of
colatitude is that one ensuring a very smooth transition to the equatorial region
and allows to reproduce qualitatively and quantitatively the latitudinal profile of the GCR intensity, and the latitudinal dip shift with respect to the ecliptic plane.
Finally we determined how the \textit{polar region} diffusion is mostly responsible of the proton intensity latitudinal
gradient observed
in the inner heliosphere with the Ulysses spacecraft during 1995.

\section*{Acknowledgements}
KK wishes to acknowledge VEGA grant agency project 2/0081/10 for support.~Finally, the authors acknowledge the use of NASA/GSFC's Space Physics Data Facility's OMNIWeb service, and OMNI data.

\section*{References}
\bibliography{biblio}

\begin{thebibliography}{10}

\bibitem{parker1965}
E.~N. Parker.
\newblock The passage of energetic charged particles through interplanetary
  space.
\newblock {\em Plan. Space Sci.}, 13:9, 1965.

\bibitem{Bobik2011ApJ}
P.~{Bobik}, G.~{Boella}, M.~J. {Boschini}, C.~{Consolandi}, S.~{Della Torre},
  M.~{Gervasi}, D.~{Grandi}, K.~{Kudela}, S.~{Pensotti}, P.~G. {Rancoita}, and
  M.~{Tacconi}.
\newblock {Systematic Investigation of Solar Modulation of Galactic Protons for
  Solar Cycle 23 Using a Monte Carlo Approach with Particle Drift Effects and
  Latitudinal Dependence}.
\newblock {\em Astrophys. J.}, 745:132, February 2012.

\bibitem{rancoita2011}
C.~{Leroy} and P.-G. {Rancoita}.
\newblock {\em {Principles of Radiation Interaction in Matter and Detection,
  3rd Edition}}.
\newblock World Scientific Publishing Co, 2011.

\bibitem{strauss2012}
R.~D. {Strauss}, M.~S. {Potgieter}, and S.~E.~S. {Ferreira}.
\newblock {Modeling ground and space based cosmic ray observations}.
\newblock {\em Adv. Space Res.}, 49:392--407, January 2012.

\bibitem{parker58}
E.~N. Parker.
\newblock Dynamics of the interplanetary gas and magnetic fields.
\newblock {\em Astrophys. J.}, 128:664, 11 1958.

\bibitem{JokipiiKota89}
J.~R. Jokipii and J.~Kota.
\newblock The polar heliospheric magnetic field.
\newblock {\em Geophys. Res. Lett.}, 16:1--4, Jan 1989.

\bibitem{langner2004}
U.W. Langner.
\newblock {\em Effect of termination shock acceleraion on cosmic ray in the
  helisphere}.
\newblock PhD thesis, Potchestroom University, Potchestroom, 2004.

\bibitem{Heber1996}
B.~{Heber}, W.~{Droege}, P.~{Ferrando}, L.~J. {Haasbroek}, H.~{Kunow},
  R.~{Mueller-Mellin}, C.~{Paizis}, M.~S. {Potgieter}, A.~{Raviart}, and
  G.~{Wibberenz}.
\newblock {Spatial variation of {$>$}40MeV/n nuclei fluxes observed during the
  ULYSSES rapid latitude scan.}
\newblock {\em Astron. Astrophys.}, 316:538--546, December 1996.

\bibitem{Simpson1996a}
J.~A. {Simpson}.
\newblock {Ulysses cosmic-ray investigations extending from the south to the
  north polar regions of the Sun and heliosphere.}
\newblock {\em Nuovo Cimento C}, 19:935--943, December 1996.

\bibitem{Strauss2011}
R.~D. {Strauss}, M.~S. {Potgieter}, I.~{B{\"u}sching}, and A.~{Kopp}.
\newblock {Modeling the Modulation of Galactic and Jovian Electrons by
  Stochastic Processes}.
\newblock {\em Astrophys. J.}, 735:83, July 2011.

\bibitem{parker1957}
E.~N. {Parker}.
\newblock {Newtonian Development of the Dynamical Properties of Ionized Gases
  of Low Density}.
\newblock {\em Phys. Rev.}, 107:924--933, August 1957.

\bibitem{Parker1960}
E.~N. {Parker}.
\newblock {The Hydrodynamic Theory of Solar Corpuscular Radiation and Stellar
  Winds.}
\newblock {\em Astrophys. J.}, 132:821, November 1960.

\bibitem{Parker1961a}
E.~N. {Parker}.
\newblock {Sudden Expansion of the Corona Following a Large Solar Flare and the
  Attendant Magnetic Field and Cosmic-Ray Effects.}
\newblock {\em Astrophys. J.}, 133:1014, May 1961.

\bibitem{Parker1963}
E.~N. {Parker}.
\newblock {\em {Interplanetary dynamical processes.}}
\newblock New York, Interscience Publishers, 1963., 1963.

\bibitem{Hattingh1995}
M.~Hattingh and R.~A. Burger.
\newblock A new simulated wavy neutral sheet drift model.
\newblock {\em Adv. Space Res.}, 16(9):213--216, 1995.

\bibitem{SW_web}
J.~H. {King} and N.~E. {Papitashvili}.
\newblock {Solar wind spatial scales in and comparisons of hourly Wind and ACE
  plasma and magnetic field data}.
\newblock {\em J. Geophys. Res.-Space}, 110(A9):A02104, February 2005.

\bibitem{OMNIWeb}
NASA-OMNIweb.
\newblock online database http://omniweb.gsfc.nasa.gov/form/dx1.html, 2012.

\bibitem{JokipiiThomas1981}
J.~R. {Jokipii} and B.~{Thomas}.
\newblock {Effects of drift on the transport of cosmic rays. IV - Modulation by
  a wavy interplanetary current sheet}.
\newblock {\em Astrophys. J.}, 243:1115--1122, February 1981.

\bibitem{Jokipii77}
J.~R. {Jokipii}, E.~H. {Levy}, and W.~B. {Hubbard}.
\newblock Effect of particle drift on cosmic-ray transport. i. general
  properties, application to solar modulation.
\newblock {\em Astrophys. J. Lett.}, 213:L85--L88, April 1977.

\bibitem{Marsch2003}
E.~{Marsch}, W.~I. {Axford}, and J.~F. {McKenzie}.
\newblock {Solar wind}.
\newblock In B.~N. {Dwivedi}, editor, {\em Dynamic Sun}, pages 374--402, 2003.

\bibitem{JokipiLev1977}
J.~R. {Jokipii} and E.~H. {Levy}.
\newblock {Effects of particle drifts on the solar modulation of galactic
  cosmic rays}.
\newblock {\em Astrophys. J. Lett.}, 213:L85--L88, April 1977.

\bibitem{Potgieter85}
M.~S. Potgieter and H.~Moraal.
\newblock A drift model for the modulation of galactic cosmic rays.
\newblock {\em Astrophys. J.}, 294(part 1):425--440, 1985.

\bibitem{BurgerHatting1995}
R.~A. {Burger} and M.~{Hattingh}.
\newblock {Steady-State Drift-Dominated Modulation Models for Galactic Cosmic
  Rays}.
\newblock {\em Astroph. and Sp. Sc.}, 230:375--382, August 1995.

\bibitem{Jokipii1970}
J.~R. {Jokipii} and E.~N. {Parker}.
\newblock {On the Convection, Diffusion, and Adiabatic Deceleration of Cosmic
  Rays in the Solar Wind}.
\newblock {\em Astroph. J.}, 160:735, May 1970.

\bibitem{FFM1968}
L.~J. {Gleeson} and W.~I. {Axford}.
\newblock {Solar Modulation of Galactic Cosmic Rays}.
\newblock {\em Astrophys. J.}, 154:1011, December 1968.

\bibitem{Yamada1998}
Y.~{Yamada}, S.~{Yanagita}, and T.~{Yoshida}.
\newblock {A stochastic view of the solar modulation phenomena of cosmic rays}.
\newblock {\em Geophys. Res. Lett.}, 25:2353--2356, July 1998.

\bibitem{GervasiEtAl1999}
M.~{Gervasi}, P.~G. {Rancoita}, I.~G. {Usoskin}, and G.~A. {Kovaltsov}.
\newblock {Monte-Carlo approach to Galactic Cosmic Ray propagation in the
  Heliosphere}.
\newblock {\em Nucl. Phys. B - Proc. Sup.}, 78:26--31, August 1999.

\bibitem{Zhang1999}
M.~{Zhang}.
\newblock {A Markov Stochastic Process Theory of Cosmic-Ray Modulation}.
\newblock {\em Astrophys. J.}, 513:409--420, March 1999.

\bibitem{alanko2007}
K.~{Alanko-Huotari}, I.~G. {Usoskin}, K.~{Mursula}, and G.~A. {Kovaltsov}.
\newblock {Stochastic simulation of cosmic ray modulation including a wavy
  heliospheric current sheet}.
\newblock {\em J. Geophys. Res.}, 112:A08101, 2007.

\bibitem{PeiBurger2010}
C.~{Pei}, J.~W. {Bieber}, R.~A. {Burger}, and J.~{Clem}.
\newblock {A general time-dependent stochastic method for solving Parker's
  transport equation in spherical coordinates}.
\newblock {\em J. Geophys. Res.-Space}, 115(A12107), December 2010.

\bibitem{Gardiner1985}
C.W. Gardiner.
\newblock {\em Handbook of stochastic methods: for physics, chemistry and
  natural sciences}.
\newblock Springer Edition, 1985.

\bibitem{jokipii1971}
J.~R. {Jokipii}.
\newblock {Propagation of cosmic rays in the solar wind.}
\newblock {\em Rev. Geoph. Space Phys.}, 9:27--87, 1971.

\bibitem{Palmer1982}
I.~D. {Palmer}.
\newblock {Transport coefficients of low-energy cosmic rays in interplanetary
  space}.
\newblock {\em Rev. Geophys. Space Ge.}, 20:335--351, May 1982.

\bibitem{PotgieterFerreira2002}
M.~S. {Potgieter} and S.~E.~S. {Ferreira}.
\newblock {Effects of the solar wind termination shock on the modulation of
  Jovian and galactic electrons in the heliosphere}.
\newblock {\em J. Geophys. Res.-Space}, 107:1089, July 2002.

\bibitem{droge2005}
W.~{Dr{\"o}ge}.
\newblock {Probing heliospheric diffusion coefficients with solar energetic
  particles}.
\newblock {\em Adv. Space Res.}, 35:532--542, 2005.

\bibitem{McDonald1997}
F.~B. {McDonald}, P.~{Ferrando}, B.~{Heber}, H.~{Kunow}, R.~{McGuire},
  R.~{M{\"u}ller-Mellin}, C.~{Paizis}, A.~{Raviart}, and G.~{Wibberenz}.
\newblock {A comparative study of cosmic ray radial and latitudinal gradients
  in the inner and outer heliosphere}.
\newblock {\em J. Geophys. Res.}, 102:4643--4652, March 1997.

\bibitem{SSN}
{SIDC-team}.
\newblock {The International Sunspot Number}.
\newblock {\em {Monthly Report on the International Sunspot Number, online
  catalogue}}, 1964-2010.

\bibitem{usoskin2011}
I.~G. {Usoskin}, G.~A. {Bazilevskaya}, and G.~A. {Kovaltsov}.
\newblock {Solar modulation parameter for cosmic rays since 1936 reconstructed
  from ground-based neutron monitors and ionization chambers}.
\newblock {\em J. Geophys. Res.-Space}, 116:A02104, February 2011.

\bibitem{BaloghEtAl1995}
A.~{Balogh}, E.~J. {Smith}, B.~T. {Tsurutani}, D.~J. {Southwood}, R.~J.
  {Forsyth}, and T.~S. {Horbury}.
\newblock {The Heliospheric Magnetic Field Over the South Polar Region of the
  Sun}.
\newblock {\em Science}, 268:1007--1010, May 1995.

\bibitem{Moraal1990}
H.~{Moraal}.
\newblock {Proton Modulation Near Solar Minimim Periods in Consecutive Solar
  Cycles}.
\newblock In {\em International Cosmic Ray Conference}, volume~6, page 140,
  1990.

\bibitem{SmithBieber1991}
C.~W. {Smith} and J.~W. {Bieber}.
\newblock {Solar cycle variation of the interplanetary magnetic field spiral}.
\newblock {\em Astroph. J.}, 370:435--441, March 1991.

\bibitem{Fisk1996}
L.~A. {Fisk}.
\newblock {Motion of the footpoints of heliospheric magnetic field lines at the
  Sun: Implications for recurrent energetic particle events at high
  heliographic latitudes}.
\newblock {\em J. Geophys. Res.}, 101:15547--15554, July 1996.

\bibitem{HitgeBurger2010}
M.~{Hitge} and R.~A. {Burger}.
\newblock {Cosmic ray modulation with a Fisk-type heliospheric magnetic field
  and a latitude-dependent solar wind speed}.
\newblock {\em Adv. Space Res.}, 45:18--27, January 2010.

\bibitem{Sandersonetal1995}
T.~R. {Sanderson}, R.~G. {Marsden}, K.-P. {Wenzel}, A.~{Balogh}, R.~J.
  {Forsyth}, and B.~E. {Goldstein}.
\newblock {High-Latitude Observations of Energetic Ions During the First
  ULYSSES Polar Pass}.
\newblock {\em Space Sci. Rev.}, 72:291--296, April 1995.

\bibitem{Marsden2001}
R.~G. {Marsden}.
\newblock {The 3-D Heliosphere at Solar Maximum}.
\newblock {\em The Publications of the Astronomical Society of the Pacific},
  113:129--130, January 2001.

\bibitem{BaloghetAl2001}
A.~{Balogh}, R.~G. {Marsden}, and E.~J. {Smith}.
\newblock {\em {The heliosphere near solar minimum. The Ulysses perspective}}.
\newblock Springer-Praxis Books in Astrophysics and Astronomy, 2001.

\bibitem{Heber1997}
B.~{Heber}, M.~S. {Potgieter}, and P.~{Ferrando}.
\newblock {Solar modulation of galactic cosmic rays: the 3D heliosphere}.
\newblock {\em Adv. Space Res.}, 19:795--804, May 1997.

\bibitem{potgieter2000}
M.~S. {Potgieter}.
\newblock {Heliospheric modulation of cosmic ray protons: Role of enhanced
  perpendicular diffusion during periods of minimum solar modulation}.
\newblock {\em J. Geophys. Res.}, 105:18295--18304, 2000.

\bibitem{Hoeksema1995}
J.~T. {Hoeksema}.
\newblock {The Large-Scale Structure of the Heliospheric Current Sheet During
  the ULYSSES Epoch}.
\newblock {\em Space Sci. Rev.}, 72:137--148, April 1995.

\bibitem{Ferreira2004}
S.~E.~S. {Ferreira} and M.~S. {Potgieter}.
\newblock {Long-Term Cosmic-Ray Modulation in the Heliosphere}.
\newblock {\em Astrophys. J.}, 603:744--752, March 2004.

\bibitem{bess_prot}
Y.~{Shikaze}, S.~{Orito}, T.~{Mitsui}, and {BESS Collaboration}.
\newblock {Measurements of 0.2 20 GeV/n cosmic-ray proton and helium spectra
  from 1997 through 2002 with the BESS spectrometer}.
\newblock {\em Astrop. Phys.}, 28:154--167, 2007.

\bibitem{AMS01_prot}
M.~{Aguilar}, J.~{Alcaraz}, J.~{Allaby}, and {AMS Collaboration}.
\newblock {The Alpha Magnetic Spectrometer (AMS) on the International Space
  Station: Part I - results from the test flight on the space shuttle}.
\newblock {\em Phys. Rep.}, 366:331--405, 2002.

\bibitem{Pamela_Prot}
O.~{Adriani}, G.~C. {Barbarino}, G.~A. {Bazilevskaya}, and {PAMELA
  Collaboration}.
\newblock {PAMELA Measurements of Cosmic-Ray Proton and Helium Spectra}.
\newblock {\em Science}, 332:69, 2011.

\bibitem{Ndiitwani2005}
D.~C. {Ndiitwani}, S.~E.~S. {Ferreira}, M.~S. {Potgieter}, and B.~{Heber}.
\newblock {Modelling cosmic ray intensities along the Ulysses trajectory}.
\newblock {\em Annales Geophysicae}, 23:1061--1070, March 2005.

\end{thebibliography}

\end{document}